\documentclass[twocolumn,amsmath,amssymb]{revtex4}

\usepackage[dvips]{graphicx,color}
\usepackage{dcolumn} 
\usepackage{bm}

\begin{document}

\title{CaCu$_3$Ti$_4$O$_{12}$/CaTiO$_3$ Composite Dielectrics: 
A Ba/Pb-free Ceramics with High Dielectric Constants}

\author{W. Kobayashi}
\email{kobayashi-wataru@suou.waseda.jp}
\affiliation{Department of Applied Physics, Waseda University, 
Tokyo 169-8555, Japan}

\author{I. Terasaki}
\affiliation{Department of Applied Physics, Waseda University, 
Tokyo 169-8555, Japan}

\date{\today}

\begin{abstract}

We have measured dielectric properties of Ca$_{1+x}$Cu$_{3-x}$Ti$_4$O$_{12}$ 
($x$ = 0, 0.1, 0.5, 1, 1.5, 2, 2.9 and 3), and have found 
that Ca$_2$Cu$_2$Ti$_4$O$_{12}$ (a composite of CaCu$_3$Ti$_4$O$_{12}$ and 
CaTiO$_3$) exhibits a high dielectric constant of 1800 with 
a low dissipation factor of 0.02 below 100 kHz from 220 to 300 K. 
These are comparable to (or even better than) those of the Pb/Ba-based ceramics,
which could be attributed to a barrier layer of CaTiO$_3$
on the surface of the CaCu$_3$Ti$_4$O$_{12}$ grains. 
The composite dielectric ceramics reported here are environmentally benign as 
they do not contain Ba/Pb.

\end{abstract}
\maketitle

It is widely known that high-dielectric ceramic capacitors 
primarily composed of Ba/Pb-based perovskite oxides are indispensable 
to modern electronic devices. 
These oxides have made a great contribution to development of 
electronics since their discoveries in forties$-$fifties \cite{1, 2}. 
As seen in a recent discovery of Pb-free piezoceramics by Saito {\it et al.} \cite{3}, 
there increase pressing needs to use environmentally-friendly materials. 
Ceramic capacitors should also be replaced with Pb/Ba-free ones, if possible. 

One can imagine that a good ceramic capacitor should have a large 
dielectric constant $\varepsilon'$. This is, however, not a sufficient condition; 
it should also require a low dielectric loss $\varepsilon''$ (or a low dissipation 
factor tan$\delta$ = $\varepsilon''$/$\varepsilon'$, which is related to loss of stored electricity), 
and a flat temperature dependence of $\varepsilon'$. At present, Ba/Pb-based 
perovskite oxides (e.g. BaTiO$_3$ and Pb(Mg$_{1/3}$Nb$_{2/3}$)O$_3$ ) are widely 
used as capacitor materials, and exhibit a high $\varepsilon'$ (1000-20000) 
with a low tan$\delta$ (0.01-0.2). Unfortunately, $\varepsilon'$ of BaTiO$_3$ changes 
rapidly around the ferroelectric transition temperature (278 K and 400 K), 
which does not satisfy the weak temperature dependence near 
room temperature. Although Pb(Mg$_{1/3}$Nb$_{2/3}$)O$_3$ does not show 
clear ferroelectric transition, $\varepsilon'$ of Pb(Mg$_{1/3}$Nb$_{2/3}$)O$_3$ also 
changes rapidly around 250 K. To reduce the rapid change, 
various composites with other materials (e.g. CaTiO$_3$, PbTiO$_3$, 
BaZrO$_3$ and PbZrO$_3$) have been examined \cite{4, 5}.

Subramanian {\it et al.} \cite{6} discovered that ceramic samples of CaCu$_3$Ti$_4$O$_{12}$ have 
a very large $\varepsilon'$ of 10000, which is nearly constant over a wide temperature 
range from 100 to 400 K below 1 MHz \cite{7}. This is quite promising for ceramic 
capacitors, and it has attracted many researchers' interests from both 
technological \cite{8, 9} and scientific \cite{10, 11, 12, 13} points of view. 
However, CaCu$_3$Ti$_4$O$_{12}$ has not yet been applied 
because of the large tan$\delta$ of 0.15 around room temperature.

\begin{figure}[tb]
\begin{center}
\includegraphics[width=8cm,clip]{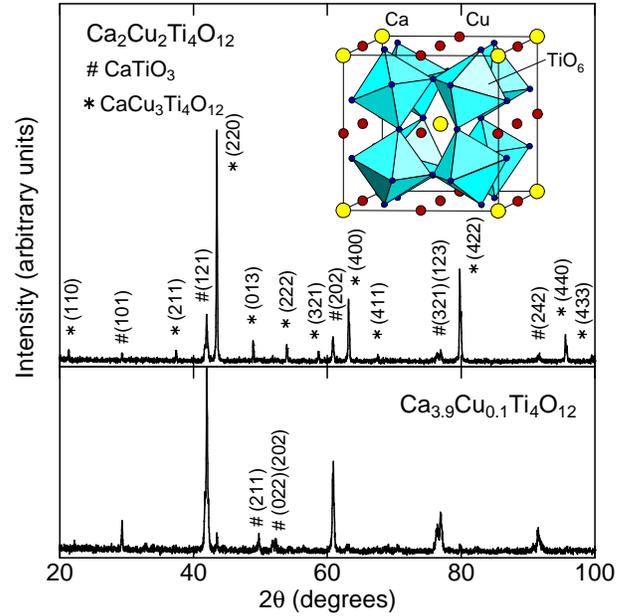}
\end{center}
\caption{X-ray diffraction patterns of Ca$_2$Cu$_2$Ti$_4$O$_{12}$ 
and Ca$_{3.9}$Cu$_{0.1}$Ti$_4$O$_{12}$ 
at room temperature. Inset shows the crystal structure of CaCu$_3$Ti$_4$O$_{12}$.}
\label{f1}
\end{figure} 

In a previous work \cite{13}, we reported remarkable substitution effects in CaCu$_3$Ti$_4$O$_{12}$, 
where only 2 \% substitution of Mn for Cu reduces $\varepsilon'$ from 10000 down to 100. 
We further attempted to improve the dielectric properties of CaCu$_3$Ti$_4$O$_{12}$ by 
partial substitution for Ca, Cu and Ti. Though some of the solid solutions 
exhibited higher $\varepsilon'$, almost all the compounds showed poorer dielectric 
performance because of their high tan$\delta$. 
Thus, as a next move, we expected that a certain kind of two-phase 
composite could improve the dielectric properties. Eventually, we have 
found that a composite of CaCu$_3$Ti$_4$O$_{12}$ (high $\varepsilon'$) and CaTiO$_3$ (low tan$\delta$) 
yields high dielectric performance. Here, we present that a composite 
of CaCu$_3$Ti$_4$O$_{12}$ : CaTiO$_3$ = 2 : 1 
(the nomial composition is Ca$_2$Cu$_2$Ti$_4$O$_{12}$)
exhibits a high $\varepsilon'$ (1800) with a low tan$\delta$ ($\leq  
0.02$) below 100 kHz, which is nearly constant from 220 to 300 K.

\begin{figure}[tb]
\begin{center}
\includegraphics[width=8cm,clip]{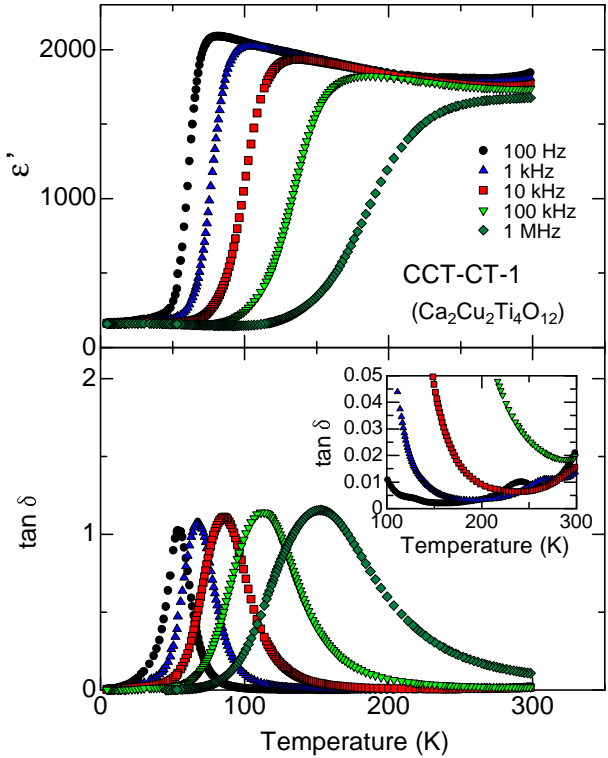}
\end{center}
\caption{Dielectric constant and dissipation factor of CCT-CT-1.}
\label{f2}
\end{figure} 

\begin{figure}[t]
\begin{center}
\includegraphics[width=8cm,clip]{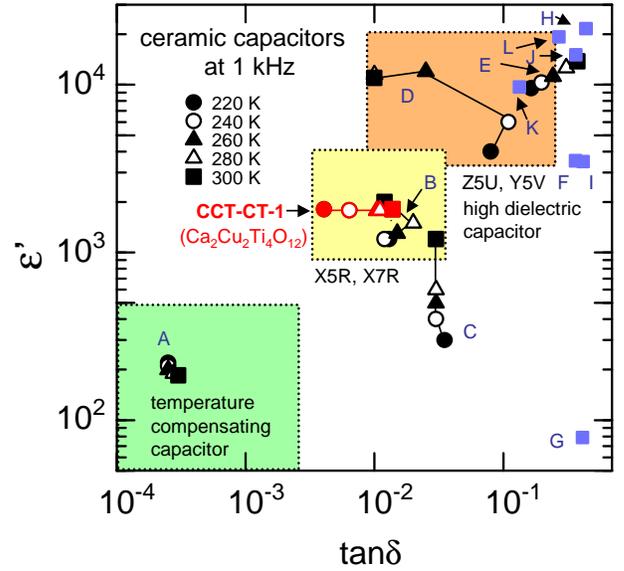}
\end{center}
\caption{Dielectric properties of some famous capacitor materials. 
Dotted rectangles represent roughly classified categories of dielectric materials. 
Temperature compensating capacitor is located in the area where $10<$ $\varepsilon'$ $<500$, 
$\mid$$\Delta$$\varepsilon'$$\mid$  $\leq$  5 \% and $10^{-4}<$ tan$\delta$ $<10^{-3}$. 
Here $\Delta$$\varepsilon'$ is defined as ($\varepsilon'$$_T$$-$$\varepsilon'$$_{20}$) / 
$\varepsilon'$$_{20}$$\times$100 [\%], 
where $\varepsilon'$$_{20}$ and $\varepsilon'$$_T$ represent $\varepsilon'$ 
at 20 $^{\circ}$C and $\varepsilon'$ at T $^{\circ}$C, respectively. 
X5R [X7R] is defined as a capacitor that exhibits $\mid$$\Delta$$\varepsilon'$$\mid$  $\leq$  15 \% 
in the temperature range from -55 $^{\circ}$C (218 K) to +85 $^{\circ}$C (358 K) 
[from -55 $^{\circ}$C (218 K) to +125 $^{\circ}$C (398 K)]. 
Z5U is defined as a capacitor that exhibits -55 \% $\leq$ $\Delta$$\varepsilon'$ 
$\leq$ +22 \% in the temperature range from +10 $^{\circ}$C (283 K) to +85 $^{\circ}$C (358 K). 
Y5V is defined as a capacitor that exhibits -82 \% $\leq$ $\Delta$$\varepsilon'$ 
$\leq$ +22 \% in the temperature range from -30 $^{\circ}$C (243 K) to +85 $^{\circ}$C (358 K). 
Alphabetical letters correspond to dielectric materials; A: CaTiO$_3$, B: BaTiO$_3$, 
C: Pb(Sc$_{1/2}$Ta$_{1/2}$)O$_3$, D: Pb(Mg$_{1/3}$Nb$_{2/3}$)O$_3$, 
E: CaCu$_3$Ti$_4$O$_{12}$, F: CaCu$_{2.97}$Mn$_{0.03}$Ti$_4$O$_{12}$, 
G: CaCu$_{2.94}$Mn$_{0.06}$Ti$_4$O$_{12}$, H: Ca$_{0.95}$Na$_{0.05}$Cu$_3$Ti$_4$O$_{12}$, 
I: Ca$_{0.95}$La$_{0.05}$Cu$_3$Ti$_4$O$_{12}$, J: Ca$_{0.95}$Sr$_{0.05}$Cu$_3$Ti$_4$O$_{12}$, 
K: CaCu$_3$Ti$_{3.8}$Sn$_{0.2}$O$_{12}$ and L: CaCu$_{2.85}$Zn$_{0.15}$Ti$_4$O$_{12}$. 
The data from A to D are taken from ref. \cite{4}.}
\label{f3}
\end{figure} 

Polycrystalline samples of Ca$_{1+x}$Cu$_{3-x}$Ti$_4$O$_{12}$ ($x$ = 0, 0.1, 0.5, 1, 1.5, 2, 2.9 and 3) 
were prepared by a solid-state reaction. Stoichiometric amounts of CaCO$_3$, CuO 
and TiO$_2$ were mixed, and were calcined at 1050 $^{\circ}$C for 12 h in air. 
The product was finely ground, pressed into a pellet, and sintered at 1090 $^{\circ}$C 
for 24 h in air. The pellet was well prepared with a density of nearly 100 \% 
of the theoretical one. The X-ray diffraction (XRD) of the sample was measured 
using a standard diffractometer (Rint-2000, Rigaku) with Fe K$\alpha$ radiation 
as the X-ray source in the $\theta$-2$\theta$  scan mode. Dielectric constant and 
dissipation factor of the sample with a couple of Ag-paste electrodes 
(DUPONT 4922N) were measured with a parallel-plate capacitor arrangement 
using an AC four-probe method with an LCR meter (Agilent-4284A) from 
10$^2$ to 10$^6$ Hz. The temperature was varied between 4.2 and 300 K in a liquid He cryostat.

Figure 1 shows the X-ray diffraction patterns for $x=1$
(the nominal composition of Ca$_2$Cu$_2$Ti$_4$O$_{12}$) 
and for $x=0.1$ (the nominal composition of
Ca$_{3.9}$Cu$_{0.1}$Ti$_4$O$_{12}$), 
where all the peaks are indexed as a composite 
of CaCu$_3$Ti$_4$O$_{12}$ (CCTO) and CaTiO$_3$ (CTO). 
A single phase does not appear in the nominal composition range of 
$0.1\leq x \leq 2.9$. 
This is rather surprising, because the crystal structure of CCTO is very 
similar to that of CTO: CCTO belongs to an ordered perovskite 
where the Ca ion in CTO is replaced by Ca$_{1/4}$Cu$_{3/4}$. 
The best dielectric properties are seen in the $x=1$ sample,
and we will call it CCT-CT-1 hereafter.

Figure 2 shows the dielectric constant $\varepsilon'$ and the dissipation factor 
tan$\delta$ of CCT-CT-1. In a frequency range 
from 10$^2$ to 10$^5$ Hz and a temperature range from 220 to 300 K, a 
large $\varepsilon'$ of 1800 is almost independent of temperature. 
It should be emphasized that tan$\delta$ remains at a low value 
of less than 0.02 in the same frequency and temperature ranges. 
This value is one order of magnitude smaller than that of CCTO. 
Below about 220 K, $\varepsilon'$ rapidly decrease, and tan$\delta$ shows a large peak, 
which are explained in terms of dielectric relaxation \cite{10}.

Now we compare CCT-CT-1 with other ceramic capacitors. 
Figure 3 shows $\varepsilon'$ and tan$\delta$ for some famous ceramic materials. 
These materials are classified into two large categories according 
to values of their $\varepsilon'$ and tan$\delta$. The one category is for temperature 
compensating capacitor, and the other one is for high dielectric capacitor. 
The latter is further classified into X5R, X7R, Z5U and Y5V of 
EIA (Electronic Industries Association in USA) standards 
(For the detailed regulations, see figure caption.). 
These categories are roughly represented by dotted rectangles in Fig. 3. 
The dielectric performance of CCT-CT-1 is in the X5R/X7R standards,
which is comparable with those of Samples B (BaTiO$_3$) and C (Pb(Sc$_{1/2}$Ta$_{1/2}$)O$_3$). 
In particular, temperature variation 
of $\varepsilon'$ of CCT-CT-1 is +1.0 \%  at 1 kHz from 218 K (-55 $^{\circ}$C) 
to 300 K (+27 $^{\circ}$C), which is the smallest value in X5R/X7R, 
and is even superior to that of sample A (CTO). 
Samples F-I in Fig. 3 represent dielectric properties of 
partially substituted samples for Ca, Cu and Ti in CCTO. 
They have the large tan$\delta$ in common, which suggests that 
the good dielectric properties of CCT-CT-1 are 
unlikely to come from simple substitution effects.

\begin{figure}[t]
\begin{center}
\includegraphics[width=8cm,clip]{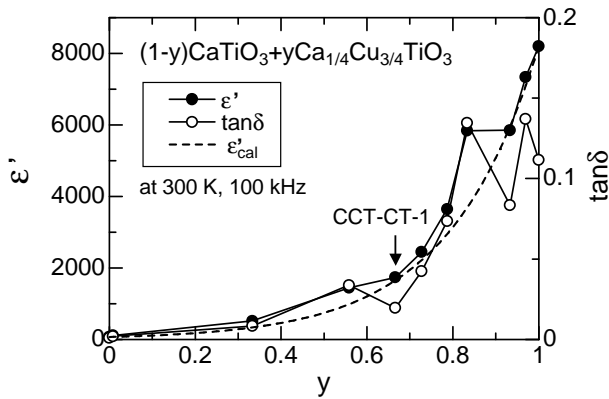}
\end{center}
\caption{Room-temperature dielectric properties of 
(1-$y$)CaTiO$_3$ + $y$Ca$_{1/4}$Cu$_{3/4}$TiO$_3$ at 100 kHz.} 
\label{f4}
\end{figure} 

Let us discuss an origin of the large $\varepsilon'$ and the low tan$\delta$ 
of CCT-CT-1. 
Figure 4 shows the room-temperature dielectric properties 
of (1-$y$)CaTiO$_3$ + $y$Ca$_{1/4}$Cu$_{3/4}$TiO$_3$ 
at 100 kHz, 
where $y$ represents a volume fraction of CCTO
($y=0.66$ corresponds to CCT-CT-1). 
$\varepsilon'_{\rm cal}$ shown by the dotted curve
is a calculation from Lichtenecker's logarithmic law
written by
\begin{equation}
\ln\varepsilon'_{\rm cal} = y\ln\varepsilon'_{\rm CCTO} + 
(1-y)\ln\varepsilon'_{\rm CTO},
\label{e1}
\end{equation}
where $\varepsilon'_{\rm CCTO}$ and $\varepsilon'_{\rm CTO}$ 
represent dielectric constants of CCTO and CTO, respectively. 
Obviously, the experimental data are in excellent agreement with $\varepsilon'_{\rm cal}$.

Equation (1) was first suggested by Lichtenecker in 1926 \cite{16}, 
and has been used by many researchers.
In fact, it well explains dielectric properties of various composites,
such as BaTiO$_3$/CaTiO$_3$ \cite{5}, BaTiO$_3$/BaTiO$_3$-gel \cite {17}, 
aluminum powder/epoxy resin \cite{18}, 
and human blood \cite{19}. 
According to Zakri et al. \cite{20}, 
the necessary conditions of Eq. (1) are:
(i) ingredients in the mixture are not dissolved in the solvent, 
and 
(ii) in a two-component case, one component is uniformly 
distributed as fine particles in the background of the other one.  
We readily see that CCT-CT-1 satisfies the first condition,
because CCTO and CTO make no solid solution  as mentioned earlier.
In a microscopic point of view, this means that Ca ions do not occupy the Cu site in CCTO,
and hence the surface of the CCTO grain should be covered with CTO in CCT-CT-1,
which suggests that CTO acts as a barrier layer.
Some systems such as BaTiO$_3$-CeO$_2$ and BaTiO$_3$-Nb$_2$O$_5$-Co$_3$O$_4$, and 
BaTiO$_3$-CdBi$_2$Nb$_2$O$_9$ obeys Eq. (1),
in which the BaTiO$_3$ grain takes the so-called ``core-shell'' structure \cite{21,22,23}.
We expect that a similar core-shell structure will exist in CCT-CT-1,
which should be verified in a further study.

Preliminarily, we took scanning-electron-microscope images
to evaluate the size and distribution of the grains in CCT-CT-1.
They showed homogeneous distribution of the CCTO grains,
which satisfies the second condition for Eq. (1).
The grain size of CCTO was $2-3$ $\mu$m, which was significantly smaller than a
typical grain size (several dozen $\mu$m) of normally prepared CCTO. 
Such a small grain size of CCTO shows a high dielectric constant (2000-2500) 
with remarkably flat temperature dependence \cite{24}, 
which can be another origin for the high performance of CCT-CT-1.

Finally we will make some comments on remaining issues.
(i) We found that CCT-CT-1 showed no piezoelectricity. 
This is reasonable, because it shows no ferroelectric transition, 
which is advantageous for reduction mechanical damages in ac operation.
(ii) A sintering temperature of CCT-CT-1 is 1090 $^{\circ}$C, which is 
significantly lower than that of BaTiO$_3$ (typically 1350 $^{\circ}$C). 
This indicates that CCT-CT-1 can save considerable costs and energies,
which is another environmentally-friendly feature.
(iii) CCT-CT-1 is highly stable. After 20000 h, 
$\varepsilon'$ and tan$\delta$ remain intact within +4 \% and +13 \%, 
respectively, which is still in the range of the X5R/X7R standards. 

In summary, we have prepared a set of ceramic samples of 
Ca$_{1+x}$Cu$_{3-x}$Ti$_4$O$_{12}$ ($x$ = 0, 0.1, 0.5, 1, 1.5, 2, 2.9 and 3),
and measured the dielectric properties.
They are found to be composites of CaCu$_3$Ti$_4$O$_{12}$ and CaTiO$_3$,
whose dielectric constant obeys Lichtenecker's logarithmic law.
This means that the CaCu$_3$Ti$_4$O$_{12}$ grains distribute 
homogeneously in the CaTiO$_3$ matrix, in which CaTiO$_3$ can act as a 
barrier layer.
For $x=1$, the dielectric constant reaches as large as 1800 with a small
loss of $\tan \delta \le $2\% below 100 kHz in a wide temperature range
from 220 to 300 K.
These are the best values among dielectric ceramics in the X5R/X7R standards.
Although the dielectric constant is highly dependent on frequency,
we hope that it will be applied in a complimentary way to existing ceramics.

We would like to thank M. Fukunaga for piezoelectricity measurement,
and Y. Uesu for useful advice including critical reading of our manuscript.

\end{document}